\documentstyle{article}    %

\catcode`@=11

\long\def\@makefntext#1{\parindent 1em\noindent
 \hbox {$^{\@thefnmark})$\kern.5em}#1}
\renewcommand{\fnum@figure}{{\small\bf Fig.\kern.2em\thefigure\kern.1em}}
\renewcommand{\fnum@table}{{\small\bf Tab.\kern.2em\thetable\kern.1em}}

\newcommand{\2}{\mbox{I{\kern-.1em}I}}

\catcode`@=12

\begin{document}

\title{Extragalactic ultra-high energy cosmic rays \\
\Large \2. Comparison with experimental data}

\author{J\"org P. Rachen$\,^1$, Todor Stanev$\,^2$, and Peter L. Biermann$\,^1$
\\[10pt]
$^1$ Max-Planck-Institut f\"ur Radioastronomie \\ Auf dem H\"ugel 69 \\
D-5300 Bonn 1, Germany \\[10pt]
$^2$ Bartol Research Institute\\ University of Delaware\\
Newark, DE 19716, USA}

\date{Accepted by Astronomy \& Astrophysics: February, 1993}

\maketitle

\begin{abstract}
The hot spots of powerful Fanaroff Riley class \2 radio galaxies and
related radio quasars can produce an ultra-high energy particle
population observable at earth.  The properties of the predicted
spectrum are a spectrum which is about $E^{-2}$ below
$0.1\,$EeV\footnotemark[1], steepening to about $E^{-2.75}$ near $1$
EeV, and cutting off due to interaction with the cosmic microwave
background after a bump just below $100$ EeV (Rachen and Biermann
1992, paper UHE-CR~I); the predicted chemical composition is a strong
dominance of protons.  Here we compare this prediction with
observational data from both the Fly's Eye and the Akeno airshower
detectors.  These experimental data are now available from the
detailed analysis of the chemical composition near to EeV energies
(Gaisser et~al.~(1992) for the Fly's Eye experiment, and from Stanev
et~al.~ (1992) for the Akeno experiment) and demonstrate that below
$1\,$EeV protons show a flatter spectrum than the overall spectrum and
thus their relative proportion increases with energy.  The comparison
of the spectral data and the flux shows that the ultra high energy
component of the cosmic rays can indeed be understood as arising from
the hot spots of powerful radiogalaxies, in flux, spectrum and
chemical composition: This result put stringent limits on the
propagation of high energy cosmic rays and thus on the properties of
the intergalactic magnetic field.
\end{abstract}

\section{Introduction}

The origin of the cosmic ray particles with energies beyond the knee
at about $5\,$PeV\footnotemark[1] is still a matter of debate, and has
recently been speculated to be extragalactic (Protheroe and Szabo
1992, Salamon and Stecker 1992).  Both papers argue that these
particles gain their energy as protons in the active cores of quasars
or other active galactic nuclei, make a transition to a neutron in a
proton-photon collision, and then escape unbound by magnetic fields.
Finally, outside the parent galaxy, these neutrons decay back to a
proton. As a consequence, both these theories predict that the cosmic
rays beyond the knee should be mostly protons. Other contender
theories put the origin of these particles i) in a galactic wind shock
(Jokipii and Morfill 1987, 1991), ii) from reacceleration of existing
energetic particles (Ip and Axford 1992), iii) from neutron stars
(Hillas 1984; M\'esc\'aros and Rudak 1991), and iv) from acceleration
by a supernova shock racing through a stellar wind (V\"olk and
Biermann 1988; Biermann 1993, paper CR I; Biermann and Cassinelli
1993, paper CR \2).  Recently (Biermann 1993, paper CR I, and Stanev
et~al.~1992, paper CR IV) we have demonstrated that the theory tracing
the cosmic ray particles beyond the knee to supernova shock waves in
stellar winds, can simultaneously explain spectrum, chemical
composition and flux from $10$ TeV\footnote{We note for reference,
that $10^{9}\,{\rm eV} = 1\,{\rm GeV}$, $10^{12}\,{\rm eV} = 1\,{\rm
TeV}$, $10^{15}\,{\rm eV} = 1\,$PeV, $10^{18} \,{\rm eV} \,=\,1
\,$EeV.} to $3\,$EeV.

Near $3$ EeV the situation becomes simpler insofar as a spectral
change suggests that the origin changes.  Since the Larmor radii of
such particles become larger than the thickness of the galactic disk,
it is likely that the origin of cosmic rays beyond several EeV is
extragalactic.  It is those particles with extremely high energies
which we wish to address in this communication.

The central thesis discussed in paper UHE-CR~I (Rachen and Biermann
1993) is that these particles originate in the hot spots of radio
galaxies (Fanaroff and Riley 1974, Miley 1980).  These hot spots are
believed to be the sites of very strong shock waves (Meisenheimer
et~al.~1989); shockwaves are believed to be the sites of particle
acceleration (Drury 1983).  Such a theory has several advantages:
\begin{enumerate}
\item First of all, it can be tested since we know the space density of radio
galaxies and their cosmological evolution (Peacock 1985), and we can
directly calculate what the integrated contribution can be.  In
addition, the radio synchrotron emission spectrum gives us the
spectral behaviour of relativistic particles.  These observed spectra
suggest particle spectra very close to $E^{-2}$.
\item We obtain i) particle energies, ii) particle flux
and iii) spectrum for reasonable assumptions for the chemical composition of
the energetic particles.  We also put a constraint on the particle propagation
through the cosmos.
\end{enumerate}
\noindent With the new data analysis of the Fly's Eye data
(Gaisser et~al.~1992) and an analysis of Akeno data (paper CR IV) we
have now information on both the chemical composition and spectrum of
the ultra-high energy cosmic rays.  Furthermore, we can use the world
data set on the cosmic rays to obtain an averaged spectrum of all
cosmic rays beyond $0.1$ EeV as a basis for the absolute normalization
of the new component.

In this paper we propose to compare theory and experimental data for
the ultra-high energy particles.  In the following we first discuss
the world set of cosmic ray data and how to obtain from them a good
average spectrum beyond the knee, then the prediction, the new
chemical composition and spectral data from Fly's Eye and Akeno, then
combine prediction and data and discuss the implications, and finally
give an outlook on what to do next.

\section{The world data set on ultra-high energy cosmic rays}

The measurement of the flux of ultra high energy cosmic rays is
difficult because of their small flux (less than $0.05$ particles per
m$^2$.ster.year above $E=10^8$ GeV). Still a relatively large
statistics has been collected during the last 20 years by four
experiments, three of which are still active (Akeno (Nagano
et~al.~1984; 1992), Yakutsk (Efimov et~al.~1990) and the Fly's Eye
(Cassiday et~al.~1990; Loh et~al.~1991)). The array at Haverah Park
(Cunningham et~al.~1980) does no longer exist. A recent presentation
and comparison of the results of the four arrays is made by Sokolsky
et al.  (1992).

The first impression of these results is that significant variations
exist in the absolute normalization and the spectral details of the
fluxes measured by different arrays (see Fig.~1). This impression is
however deceiving. The energy measurement in individual air showers is
no better than $20\%$ because of fluctuations in shower development
and very small ratio of sensitive to effective area which emphasizes
the fluctuations. The small coverage is necessary in conventional
shower arrays with effective area approaching $100$ km$^2$. Systematic
errors of the same order are not only possible but very likely. The
estimate of the primary energy relies heavily on calculations of the
shower development and estimates of the acceptance of the array, which
are very difficult and may not be exact. The absolute normalization of
the cosmic ray flux given by two different sets of detectors at Akeno
(the $1$ km$^2$ and the $20$ km$^2$ arrays) is different by about 1.5
(Nagano et~al.~1992).

The usual presentation of the flux ($E^3\,\times\,F$) exaggerates the
large differences in the normalization on the $E^3\,\times\,F$ plot.
We will attempt to look for very small systematic energy shifts that
may bring data points from different experiments in good agreement.
\begin{figure}[p]
\vspace{.9\textheight}  %
\caption[]{\small The high energy cosmic ray data, both in their
original form -- shaded -- and after shifting to a common
normalization near $1$ EeV, and averaging with weighting.  The various
symbols denote the different experiments: Akeno, circles; Yakutsk,
triangles; Fly's Eye, hexagons; Haverah Park, squares.}
\end{figure}

Apart from the differences in normalization all data sets show the
same slope ($\gamma$=3.0--3.1) in the region 3$\times$10$^{17}$ --
3$\times$10$^{18}$ eV.  We use the data in this range to normalize
experiments to each other. First an average spectrum is calculated
using the original experimental errors. Then we calculate for each
experiment the energy shift $\Delta E$ that brings the data set in
best agreement with the average spectrum. The resulting shifts are
smaller than the suspected systematic errors -- we obtain $\Delta E =
-16\%$ for Yakutsk, $-13\%$ for Akeno, $-3\%$ for Haverah Park (which
has the smallest statistical errors) and $+6\%$ for the Fly's Eye. We
apply these energy shifts to the full data sets and obtain a world
data set where the scatter is smaller than the individual error bars.
Finally we bin in logarithmically equal energy bins, combine
individual data points within the same bin and produce the average
flux shown with black squares on Figure~1. This is the flux that we
shall use to estimate the strength of different chemical components
further down.

This procedure is not unique and does not eliminate the systematic
errrors in the energy derivation. We claim, however, that it
introduces a standard systematic error, which is nearly the same for
all data sets. The energy shifts required for the averaging are
remarkably small. A better way to achieve a standard energy derivation
from different experiments is a detailed study of the experimental
algorithms (Lawrence et~al.~1991) and a comparison and possible
improvement of their theoretical bases.

\section{The prediction: Hot spots of powerful radiogalaxies}

Hot spots of powerful radio galaxies and the related radio quasars
have been argued to be sites of strong particle acceleration for some
time (Biermann and Strittmatter 1987, Meisenheimer et~al.~1989).
After some initial analytical arguments (Biermann 1990, 1991) we
demonstrated in paper UHE-CR~I, following Berezinsky and Grigor'eva
(1988), that a proper calculation of the source evolution as well as
the cosmic ray particle transport through the evolving universe is
capable to produce spectra with the following properties:
\begin{enumerate}
\item Assuming the ratio of energetic protons to electrons to be
of order $10$ and all other unknown efficiencies to average out to a
factor near unity, one obtains a flux of extragalactic cosmic rays
very close to the observed flux at $10$ EeV.
\item The spectrum is approximately the source spectrum below $0.1$ EeV,
i.e., approximately $E^{-2}$ -- known from the synchrotron radio
emission of hot spots -- then steepens to approximately $E^{-2.75}$,
and cuts off just below $100$ EeV in particle energy (Greisen 1966;
Zatsepin and Kuzmin 1966; Stecker 1968).  Depending on the source
cutoff, there is a more or less pronounced bump at particle energies
just below the cutoff.  The exact shape of the spectrum is obtained by
a proper calculation of propagation and redshift.  Already above $0.1$
EeV a slight steepening from $E^{-2}$ sets in to approximately
$E^{-2.15}$ due to galactic evolution effects.
\item The shock, which produces the acceleration is due to the violent
interaction of a jet emanating from an active nucleus with the medium
in intergalactic space, and thus is expected to have chemical
abundances between those typical for the inner parts of big elliptical
galaxies ("normal" abundances, with heavy elements enriched by a
factor up to $3$ over solar) and those typical for intercluster gas
(heavy element enrichment weak to $1/3$ solar), as known from X-ray
spectroscopy of clusters of galaxies.  This means, that protons
dominate, and Helium is of order $10\%$ in number density relative to
Hydrogen; the elements Carbon, Nitrogen, Oxygen, to Iron should be
between a maximum of $3$ times solar, and negligible.  Obviously, the
propagation of nuclei heavier than Hydrogen in the intergalactic space
can lead to spallation by photonuclear interactions with both the
infrared and the microwave background, and thus lower the observed
abundances of these nuclei.
\end{enumerate}
\noindent In Fig.~2 we present a series of model calculations for an input
spectrum of $E^{-2}$ with an intrinsic exponential cutoff at different
energies in the range about $100\,$EeV, and a Hubble constant of
$75\,\rm km\,sec^{-1}\,Mpc^{-1}$.  Such calculations involve an
estimate for the conversion of radio luminosity to jet power (Rawlings
and Saunders 1991), from jet power to energy in energetic particle
populations, and, most importantly, the ratio between the energy
density of the energetic electrons, and the energetic nuclei -- after
all, the observed synchrotron emission traces only the relativistic
electrons.  The combination of these factors we refer to as the
``fudge factor'', and since it is dominated by the nuclei to electron
ratio, it is expected to be below $10$.  Indeed, low energy fits to
the Fly's Eye data for protons (see below) imply this factor to be
near $3$, and so suggest that the proton to electron ratio is of near
$20$ or less (see UHE-CR~I).
\begin{figure}[tb]
\vspace{10cm}   %
\caption[]{\small Model calculations for the extragalactic cosmic ray
proton population observed near earth, assuming FR-\2 galaxies to be
the dominant sources. We assume an input spectrum of $E^{-2}$ with an
exponential cutoff -- the logarithm of the respective cutoff Lorentz
factor is indicated -- and a Hubble constant of $75\,\rm
km\,sec^{-1}\,Mpc^{-1}$}
\end{figure}

\section[]{The new Fly's Eye and Akeno data analysis:\\high energy protons}

We have used two independent methods to estimate the extragalactic
flux from different air shower experiments. First we fit the shower
size distributions published by the Akeno experiment (Nagano
et~al.~1984) for different zenith angles. Then we use the results of a
recent analysis (Gaisser et~al.~1992) of the cosmic ray composition
around 10$^{18}$ eV from the measurements of the Fly's Eye detector
(Cassiday et~al.~1990).

The Akeno shower array is a traditional detector that measures the
number of charged particles in the shower ($N_e$) that reach the
observation level. We use data sets for vertical ({\rm
secan}$\,\theta=1.0$, atm.~depth = 920 g/cm$^2$) and slightly inclined
({\rm secan}$\,\theta=1.2$, atm.~depth = 1104 g/cm$^2$) showers. The
fitting procedure is described in detail in paper CR IV and consists
of stepping through the energy spectrum of each chemical component
with a small logarithmic step ($10^{0.01}$) and calculating the size
of a number of showers at that energy and primary mass with the
parametrization of Gaisser (1979).  The resulting sizes at both depths
are binned with the appropriate weight, and the sum for all components
is then compared with the experimental data.

In the spectrum and composition model of CR IV all galactic nuclear
components have the same spectral index of 3.07 at energies above
$4\times10^{16}\,$eV.  The comparison with the experimental $N_e$
distribution demonstrates that galactic cosmic rays fit well the
shower size distribution up to $N_e = (1--3)\times10^7$, but are
insufficient to maintain the calculated spectrum, especially for
inclined showers, in agreement with experimental data for bigger
$N_e$. A better fit requires the introduction of a flatter cosmic ray
component at energy above 10$^{16}$ eV. The agreement is achieved by
introducing a component, consisting of pure Hydrogen, with an energy
spectrum of $(1.4\pm0.3)10^{-7}\times E^{-2} \;\rm
(cm^2\,ster\,s\,GeV)^{-1}$ at energies between $10^{16}$ and
$5\times10^{17}\,$eV and $(0.47\pm0.09)\times
E^{-2.75}\rm\;(cm^2\,ster\,s\,GeV)^{-1}$ at higher energy, where $E$
is measured in GeV. Because the spectral shape of this component is
entirely different from those of the galactic cosmic rays it is very
likely to represent an emerging extragalactic cosmic ray flux.  This
flux is shown with a box on Figure~3.

The exact shape and normalization of the extragalactic component
quoted above is not an unique solution of the experimental data fit. A
slightly steeper spectrum \hbox{($\gamma = 2.1 - 2.2$)} in the lower
energy range would produce a somewhat better fit. The absolute
normalization below and above the break at 5$\times$10$^{17}$ eV is a
result of the assumption that this flat component consist only of
Hydrogen nuclei. Since protons are more efficient in generating shower
size at the observation level, any admixture of He and heavier nuclei
would require a higher overall normalization.  The spectrum quoted
above is thus the minimum flux of extragalactic cosmic ray nuclei
necessary for a fit of the Akeno shower size spectra. As discussed in
CR IV an important source of uncertainty in the fit are the
uncertainties from the extrapolation of the particle physics input to
air shower energies.

For the second estimate we use the recent results from the analysis of
the Fly's Eye measurements of the depth of maximum ($X_{\rm max}$)
distribution in terms of cosmic ray composition. The Fly's Eye is a
different type of detector, which observes directly the longitudinal
development of air showers through the detection of the fluorescent
light from the atmospheric Nitrogen atoms, induced by the shower
charged particles. The amount of fluorescent light is proportional to
the number of charged particles after an account is made for light
scattering and absorption. The data analysis fits individual data
points (taken at various atmospheric depths) to a shower profile and
derives the depth ($X_{\rm max}$) and size ($N_{\rm max}$) of the
shower maximum. $N_{\rm max}$ is proportional to the energy of the
primary nucleus and $X_{\rm max}$ depends on the energy and mass of
the primary nucleus. The sensitivity to the primary mass comes from
the rate of energy dissipation, which is faster in showers initiated
by heavy nuclei. $X_{\rm max}$ of Fe generated showers is about 100
g/cm$^2$ shallower than that of proton generated shower of the same
energy.

The basic idea of the shower analysis is to simulate a large number of
air showers and compare the results to data.  The simulation includes
a thorough account for the light production process and the
experimental triggering and efficiency as a function of energy,
distance to the detector and angle of the detected shower. The results
of the simulation are reconstructed with the algorithm developed for
and used in the analysis of experimental data. Showers of three
composition groups (Hydrogen, CNO, and Fe) were simulated for two
different equally valid extrapolations of the properties of hadronic
interactions to 10$^9$ GeV. The composition groups are defined by the
sensitivity of the procedure to primary mass. At the energies in
question Hydrogen and Helium showers are almost indistinguishable
because these nuclei have nearly the same cross section for inelastic
interactions. All nuclei with $A$ greater than 20 are also
indistinguishable, because the mass dependent difference in $X_{\rm
max}$ in this range is significantly smaller than the current detector
resolution of 45 g/cm$^2$.  In this analsyis the whole group of nuclei
with $A\, > \, 20$ is identified as Iron.

The experimental $X_{\rm max}$ distribution is then fitted with the
simulated distributions for primary nuclei of different mass. The best
fit ($\chi^2$ minimization) determines the composition in terms of the
three distinguishable components. The procedure is repeated for two
interaction models that fit the data equally well and the following
fractions of proton showers are determined: 0.13$\pm^{0.14}_{0.05}$,
0.23$\pm^{0.15}_{0.12}$ and 0.43$\pm^{0.11}_{0.19}$ at energies of
0.38, 0.63 and 1.41 EeV, respectively.  The error bars above include
both the statistical errors from the experimental statistics and the
fitting procedure and the systematic errors from the particle physics
input in the simulation, which also make the errors very asymmetric.

The corresponding proton flux (from the average spectrum of Fig.~1) is
$(8.9\pm^{9.4}_{3.1})\times 10^{-25}$, $(3.2\pm^{2.1}_{1.7})\times
10^{-25}$ and $(5.1\pm^{1.3}_{2.2})\times 10^{-26}
\rm\;(cm^2\,s\,ster\,GeV)^{-1}$ at the three energies, respectively. These
three data points are plotted on Fig.~3 and represent the upper limits
for the extragalactic proton flux measured by the Fly's Eye. Since the
detector cannot differentiate between Hydrogen and Helium these flux
values reflect the sum of the two components.

\section{Prediction and experiment}

\begin{figure}[t]
\vspace{10cm}    %
\caption[]{\small Data and model calculations for the different components
of the UHE cosmic ray population observed near earth. The best fit
extragalactic proton component is taken from Fig.~2 and fitted to the
Fly's Eye data for the proton/He component. The difference to the
total cosmic ray spectrum gives an indication for the heavy nuclei
contribution and can be fitted by an $E^{-3.1}$ spectrum with an
exponential cutoff at $5\,$EeV. The errors of this difference
component are clearly underestimated since the theoretical
uncertainties of the extragalactic contribution are disregarded.}
\end{figure}
In Fig.~3 we now combine the best curve selected from Fig.~2, that
curve with an intrinsic cutoff of $100\,$EeV, which is nearly
$E^{-2.75}$ between $1\,$EeV and $30\,$EeV.  We selected this curve
because it fits best the experimental data simultaneously in the low
and the high energy range.  The fit is quite good, considering the
error bars on the data, fitting to about $2\sigma$ or better.

Most of the existing theoretical proposals to explain the origin of
the cosmic ray particles at energies beyond the knee can be excluded
on the basis of the presented data:
\begin{enumerate}
\item Beyond the knee the particles are dominantly heavy nuclei and
not all protons as follows from the proposals by Protheroe and Szabo
(1992), and by Salamon and Stecker (1992).
\item The galactic wind model by Jokipii and Morfill (1987) would
give normal, i.e.  close to solar, abundances for these cosmic ray
particles, as would the reacceleration model (Ip and Axford 1992)
except for the highest particle energies.  Again, the observed
abundances clearly disagree with these models.
\end{enumerate}
\noindent The difference curve ought to correspond to the galactic
contribution of heavy nuclei; this is first of all confirmed by the
Fly's Eye analysis, which does suggest, that in this particle energy
range there are very few nuclei of Carbon, Nitrogen, and Oxygen, and
many heavier nuclei.  Second, over the particle energy range which is
well measured, i.e.  below a particle energy of about $10$ EeV, the
curve is well described by a powerlaw with an exponential cutoff near
$5\,$EeV, just as suggested by the earlier cosmic ray arguments
(papers CR~I, CR~\2, and CR~IV) for the galactic contribution of
cosmic rays.  At energies above $30$ EeV the subtraction is clearly
unreliable, since the difference is between two large numbers of
similar numerical value.

We note that an incorporation of a putative infrared radiation
background would slightly lower the theoretical curve in the powerlaw
section below $0.3$ EeV particle energies, so that theory and
observation would come closer.  We also emphasize that Helium and
heavier nuclei in the extragalactic contribution would influence the
high energy behaviour near $30$ EeV, and the heavier nuclei would
probably add some flux there.

\section{Outlook}

We conclude first that a detailed check of the existing prediction
with the new data now available is successful.  The extragalactic
contribution can be readily modeled in a) flux, b) spectrum, and c)
chemical composition.

There is a second important conclusion: Strong shocks in extragalactic
jets and their associated hot spots do produce energetic nuclei.  This
is of interest, since in all arguments about Gamma-ray sources like
the quasar 3C279, Mkn 421 and the like (Hartman et~al.~1992, Punch
et~al.~1992), there is always the question whether active galactic
nuclei accelerate nuclei at all in their jets.  The successful fit
made here suggests strongly, that this is the case, and that protons
dominate the process.  This gives strong support to hadronic
interaction models to explain the gamma-ray emission from such quasars
(Mannheim and Biermann 1992; Mannheim 1993).

However, there are areas where we should improve the calculation:
\begin{enumerate}
\item Clearly, using an universal source
cutoff in particle energy in form of an exponential cutoff, is much
too simple.  The detailed shape of the cutoff of any single source
might be much softer than an exponential attached to a powerlaw, and
the cutoff energy of various sources could be quite different.  Both
effects would soften the average shape of the source cutoff spectrum.
This requires a better understanding of the source structure and shock
topology.
\item We have used straight line propagation
through the universe.  Clearly, this can only be an approximation to a
situation where particles are weakly scattered.  Our model predicts,
that we see particles from nearby radiogalaxies.  That we observe
these particles at the expected spectrum, implies that the mean free
path for scattering is of order $1/3$ of the typical distance to
nearby powerful radiogalaxies, or larger, for all particle energies
above $0.1$ EeV.  This mean free path is of an order similar to the
bubble size seen in the galaxy distribution, of order $50$ Mpc (e.g.
Einasto et~al.~1989).
\item Helium and heavier nuclei are likely to become important near the high
energy cutoff of the Hydrogen population.  Here, photonuclear
interactions have to be taken into account, and this is one of the
next steps we plan to undertake.  Limits may be possible from future
data on the high energy photon background.
\item The infrared background can be estimated from the
cosmological evolution of galaxies, using deep CCD and HST data as
well as the IRAS counts.  Such estimates can also be limited by
determing the microwave background fluctuations at redshifted FIR
frequencies corresponding to early galaxy formation and evolution,
observable now near $1$ mm wavelength.
\item An independent estimate can be made for the extragalactic cosmic ray
contribution from normal and starburst galaxies, using all the same
models as discussed in the paragraph immediately above.  Normal
galaxies also produce cosmic rays, which leak out to intergalactic
space and can add up to appreciable fluxes over cosmological time
scales (see, e.g., Biermann 1991).  Such a calculation should be done
properly.
\end{enumerate}

\subsection*{Acknowledgements}
We wish to thank Dr.~T.K.~Gaisser for many discussions on the origin
of high energy cosmic rays and their propagation.  High energy
astrophysics with PLB is supported by the DFG (Bi 191/6,7,9) and the
BMFT (DARA FKZ 50 OR 9202).  Work by TS is supported by grants
NSF/PHY/8915189 and NAG5/1573. This work was supported by a NATO
travel grant to PLB and TS.

\subsection*{References}
\frenchspacing
\begin{small}
\begin{list}{}
{\labelwidth=\parindent \leftmargin=\parindent \labelsep0pt
\parsep0pt \itemsep0pt plus 2pt}
\item[Berezinsky V.S., Grigor'eva S.I.], 1988, A\&A 199, 1
\item[Biermann P.L., Strittmatter P.A.], 1987, ApJ 322, 643
\item[Biermann P.L.], 1990, in: Fazio G.G., Silberberg R.
(eds.), {\em Currents in Astrophysics and Cosmology}, \\
Cambridge University Press, publication expected in June 1993
\item[Biermann P.L.], 1991, in Nagano and Takahara 1991, p.~301
\item[Biermann P.L.], 1993, A\&A (paper CR I, in press)
\item[Biermann P.L., Cassinelli J.P.], 1993, A\&A (paper CR \2, submitted)
\item[Cassiday G.L. et al.],  1990, ApJ 356, 669
\item[Cunningham G. et al.], 1980, ApJ Lett.~236, L71
\item[Drury L.O'C.], 1983, Rep. Prog. Phys. 46, 973
\item[Efimov N.N. et al.], 1991, in Nagano and Takahara 1991, p.~20
\item[Einasto J., Einasto M., Gramann M.], 1989, MNRAS 238, 155
\item[Fanaroff B.L., Riley J.M.], 1974, MNRAS 167, 31P
\item[Gaisser T.K.],  1979,  in Proc. Cosmic Ray Workshop, Salt Lake City, Utah
\item[Gaisser T.K. et al.], 1992, Phys. Rev. D (in press)
\item[Greisen K.], 1966, Phys. Rev. Lett. 16, 748
\item[Hartman R.C. et al.],  1992, ApJ Lett.~385, L1
\item[Hillas A.M.], 1984, ARA\&A 22, 425
\item[Ip W.H., Axford W.I.],  1992, in Conf. ``Particle acceleration in cosmic
     	plasmas'', \\Eds. G.P.~Zank, T.K.~Gaisser, AIP Conf.Proc. No. 264, p.~400
\item[Jokipii J.R., Morfill G.],  1987, ApJ 312, 170
\item[Jokipii J.R., Morfill G.],  1991, in Nagano and Takahara 1991, p.~261
\item[Lawrence M.A., Reid R.J.O., Watson A.A.], 1991, J. Phys. G: Nucl. Part.
     	Phys. 17, 733
\item[Loh E.C. et al.], 1991, in Nagano and Takahara 1991, p. 345
\item[Mannheim K., Biermann P.L.], 1992, A\&A 253, L21
\item[Mannheim K.], 1993, A\&A (in press)
\item[Meisenheimer K. et al.], 1989, A\&A 219, 63
\item[M\'esc\'aros P., Rudak B.], 1991, in Nagano and Takahara, p. 311
\item[Miley G.], 1980, ARA\&A 18, 165
\item[Nagano M., et al.], 1984, J. Phys. G: Nucl. Part. Phys. 10, 1295
\item[Nagano M., Takahara F. (eds)], 1991,   Kofu-conference proceedings \\
     	``Astrophysical aspects of the most energetic cosmic rays'', World
     	Scientific, Singapore.
\item[Nagano M. et al.], 1992, J. Phys. G: Nucl. Part. Phys. 18, 423
\item[Peacock J.A.], 1985,  MNRAS 217, 601
\item[Protheroe R.J., Szabo A.P.], 1992, Phys. Rev. Lett 69, 2885
\item[Punch M. et al.], 1992, Nature 358, 477
\item[Rachen J.P., Biermann P.L.], 1993, A\&A (paper UHE-CR~I, in~press)
\item[Rawlings S., Saunders R.], 1991, Nature 349, 138
\item[Salamon M.H., Stecker F.W.], 1992,  Phys. Rev. Lett. (submitted)
\item[Sokolsky P. et al.], 1992, Phys. Reports, 217
\item[Stanev T., Biermann P.L., Gaisser T.K.], 1992, A\&A (paper CR~IV,
	submitted)
\item[Stecker F.W.], 1968,  Phys. Rev. Lett. 21, 1016
\item[V\"olk H.J., Biermann P.L.], 1988, ApJ Lett. 333, L65
\item[Zatsepin G.T., Kuzmin V.A.], 1966, JETPh Lett. 4, 78
\end{list}
\end{small}

\end{document}